# Surface state engineering of molecule-molecule interactions

Geoffrey Rojas, Xumin Chen, Donna A. Kunkel, Justin Nitz, Jie Xiao, Peter A. Dowben, Scott Simpson, Eva Zurek, and Axel Enders*

[*]   Axel Enders
Department of Physics & Astronomy
University of Nebraska-Lincoln
Lincoln, NE 68588, USA
E-mail: a.enders@me.com

Geoffrey Rojas, Xumin Chen, Donna Kunkel, Justin Nitz, Jie Xiao, Peter A. Dowben
Department of Physics & Astronomy, University of Nebraska-Lincoln
Lincoln, NE 68588, USA

Scott Simpson, Eva Zurek
Department of Chemistry, State University of New York at Buffalo
Buffalo, NY 14260-3000, USA

Abstract:

**Engineering the electronic structure of organics through interface manipulation, particularly the interface dipole and the barriers to charge carrier injection, is of essential importance to improved organic devices. This requires the meticulous fabrication of desired organic structures by precisely controlling the interactions between molecules. The well-known principles of organic coordination chemistry cannot be applied without proper consideration of extra molecular hybridization, charge transfer and dipole formation at the interfaces. Here we identify the interplay between energy level alignment, charge transfer, surface dipole and charge pillow effect and show how these effects collectively determine the net force between adsorbed porphyrin 2H-TPP on Cu(111). We show that the forces between supported porphyrins can be altered by controlling the amount of charge transferred across the interface accurately through the relative alignment of molecular electronic levels with respect to the Shockley surface state of the metal substrate, and hence govern the self-assembly of the molecules.**

The electronic properties of organics in contact with metal substrates depend on the alignment of the electronic levels and bands at the metal-organic interface and the resulting hybridization of states, as well as charge transfer to or from the adsorbate, the molecular band offsets [1, 2, 3], the emergence of interaction-induced states [4, 5], the distortion of the molecules [6] as well as changes that may occur at the substrate surface [7]. Also key to the interface electronic structure is the presence of substrate surface states [8]. Generally, the properties of metal-organic interfaces are determined by a delicate balance of competing factors and experiments usually assess only the cumulative effect of many different contributions to the interface electronic structure [6, 7]. The net charge transferred across the interface, the formation of charge dipoles, and the work function are intrinsically related effects. Often what is highlighted is the interface dipole or the work function, but the substrate surface states, a fundamental ingredient to the interface electronic structure is often poorly described. Here we demonstrate the importance of the Shockley surface states [9] in establishing the interface electronic structure using the example of tetraphenyl porphyrins (2H-TPP) chemisorbed on Cu(111). The surface state interactions with the adsorbed molecular layers are important for the charge transfer between the substrate and the molecule and the resulting surface dipoles that ultimately strongly influence the intermolecular lateral interactions. The surface state can be shifted in energy by using Ag buffer layers of varied thickness on Cu(111), thereby determining the overlap of molecular levels with substrate surface metal bands [2], the amount of charge transferred, and consequently the intermolecular forces. We can relate our findings to the observed strong repulsive intermolecular Coulomb forces and the repression of molecular self-assembly. We show that the molecule-molecule interactions can be changed from repulsive to attractive by controlling the amount of charge transferred across the interface through surface state engineering using Ag buffer layers on the Cu(111).

The 2D character of an adsorbed monolayer of 2H-TPP has been exploited for a comparative study of the occupied and unoccupied band structure of large ensembles with direct and inverse photoelectron spectroscopy (UPS and IPES), as well as of selected individuals with the tip of a scanning tunneling microscope (STM) in the local spectroscopy mode (STS). By this combination of local and area-integrating complementary methods the atomistic basis of observed features in the electronic structure became evident. STM images, taken at sub-monolayer to monolayer coverage of 2H-TPP on Cu(111), are shown in figure 1. A coverage of $\theta = 1$ ML is defined here as the maximum observed packing density within the first layer of 0.42 molecules per nm$^2$. This packing is 20% smaller, expressed in terms of areal density, than that observed on Ag(111), see figure 1e [10, 11]. The mobility of the molecules is sufficiently high for surface diffusion, as concluded from visible substrate step edge decoration (not shown), however, no nucleation is observed. The molecules remain isolated and roughly equally spaced on the terraces of the Cu(111) (figures 1b, 1c). They appear to be aligned along the three ⟨111⟩ crystallographic directions of the surface, concluded from the observed angles of multiples of 120° between the major axes of any two molecules. It can be seen by comparing Fig 1b and c that molecules are added to the first monolayer even if the gaps between the molecules are significantly smaller than the size of the molecules itself. This requires rearrangement of all molecules in the layer during deposition. Self-organization of the 2H-TPPs into networks, as found for the same molecules on Ag(111) in figure 1a and Au(111) [10, 11, 12], was not observed on Cu(111) at any coverage and sample temperature in the rage between 77 K and 500 K. We do observe, however, by inspection of figure 1(c, d) that an alignment of the molecules with respect to each other sets in as the areal density of the molecules increases. Upon reaching saturation coverage within the first layer, molecules nucleate into islands on top of the first layer. The architecture of this arrangement is a porous 2D network apparently dominated by π–π bonds, and is a different architecture than the densely packed arrangement observed for the same molecules on Ag(111) in figure 1a. We conclude from these observations that the net force between the molecules within the first layer is repulsive, while it is attractive for the molecules within the second layer.

The occupied and unoccupied electronic structure of the adsorbate-substrate system has been studied in detail with tunneling spectroscopy and combined photoemission and inverse photoemission spectroscopies, as seen in figure 2. The combined photoemission spectra of the 2H-TPP covered Cu(111) shows characteristic peaks that are not observable in the spectra of the pristine Cu(111). One feature, at +2 eV, is in reasonable agreement with the lowest unoccupied molecular orbital (LUMO) of calculated and measured spectra for similar TPP systems [13]. Also the spectra of the occupied states resemble those reported for 2H-TPP adsorbed on other noble-metal systems [14], with the highest occupied molecular orbital (HOMO) at approximately -2 eV. Within this HOMO-LUMO gap we observe an additional characteristic peak at +0.65 eV at sub-monolayer coverage, which is observed to decay rapidly in intensity with increasing coverage and is not apparent in the spectra at 3 ML coverage or more.

Complementary to the combined photoemission and inverse photoemission spectroscopy measurements, point spectroscopy measurements have been taken locally with STS over a similar energy range, see bottom panel in figure 2. Single point dI/dV spectra were taken over the molecules themselves, as well as the surrounding Cu surface at successively increasing distance from the molecule center. The observed HOMO and LUMO of the molecules are aligned well with those observed using photoelectron spectroscopy; the LUMO is seen at approx. 1.5 eV above the Fermi level, contained with the LUMO + 1 peak. The spectra taken on the bare Cu show the well-known Shockley surface state at -0.4 eV [15, 16], which is not resolved in the photoelectron spectra. This surface state is suppressed on the Cu surface covered with $\theta \geq 0.7$ ML 2H-TPP. At lower coverage, this surface state is shifted towards the Fermi level in the direct vicinity of the molecule. The spectra shown in figure 2f was taken at a distance of 6 Å from the molecule center and shows this surface state shifted upward in energy by $\Delta E = 0.2$ eV. Also within the HOMO-LUMO gap at +0.7 eV an electronic state, already known from the IPES measurements, is observed at the molecules. This peak is only observed for spectra taken of molecules in the first monolayer. Spectra taken on molecules in the second layer do not show this substrate surface state feature, and yet the characteristic LUMO and HOMO remain undisturbed.

The molecules of the second layer appear in the STM images under the same tunneling conditions with dark center and bright phenyl arms, while in the first layer the opposite is observed, the centers are bright and the phenyl arms are dark. This change in contrast is due to an electronic level rearrangement at the interface [5]. We again exploit the local nature of tunnel spectroscopy to identify local differences in the DOS. In STS point spectra taken at the center of a molecule in the second layer the new peak at +0.65 eV, observed over the molecules in the first layer, does not appear. This allows us to attribute the physical origin of this state to the 2H-TPP/Cu interface. The electronic states in this energy range have been observed previously for other porphyrin-based surface systems on Ag(111) as well as Cu(111) with photoelectron spectroscopy [5, 17, 18], and have been heretofore ascribed to the shifted LUMO of the porphyrin macrocycle. However, the absence of the energy state at +0.65 eV in the second monolayer provides now evidence that this state is an interface state.

Measurements of the local work function, $\Phi$, have also been made using the STM. We have characterized and measured the local work function to evaluate the local surface dipoles, following a procedure similar to that published in reference [19] and described in the supplementary material. The so measured work function of the Cu(111) is $\Phi_{Cu} = (4.9 \pm 0.2)$ eV.

With 2H-TPP deposited, we find a decrease of the work function by $\Delta\Phi \sim (-2.0 \pm 0.5)$ eV over the center of TPP molecules, and an increase of $\Delta\Phi \sim +(1.0 \pm 0.4)$ eV at the boundary of the molecule's macrocycle. While these data are in quantitative agreement with the net work function shift of 0.84 eV found for 1 ML 2H-TPP on Ag(111) [5], the particular advantage of these local measurements is that they reveal a significant amount of spatial variance. For clarification of the spatial variance, a map of the work function has been measured in a square area across the molecule and its surrounding from 100 × 100 separately performed point spectra. This $\Phi$-map is shown together with an STM image of a lone 2H-TPP on Cu(111) in figure 3. By comparing both results in this manner the spatial dependence of the work function can be associated with local chemical components of the adsorbed molecule, and with the locally measured density of states. The $\Phi$ drops significantly over the location of the central pyrolines while increasing relative to the bare Cu(111) over the surrounding hydrogen edges and phenyl ligands. Surrounding the molecule in a narrow band there is a slight drop in the Cu(111) work function. This band corresponds to the area where the upward shift in the surface state was observed, too.

The electronic interactions at the 2H-TPP / Cu and the 2H-TPP / Ag interfaces can be understood using the results of density functional theory calculations, undertaken as described in the supplementary material. The computational results, summarized in Table 1, show that the binding energy of the molecules to the substrate is significantly larger on Cu(111) (3.96 eV) than on Ag(111) (0.42 eV), resulting in a much shorter distance between the molecule and the substrate and significant distortion of the molecule on Cu(111) (figure 4). In particular, the dihedral angle of the phenyl ligands changes and the ligands become nearly planar to the surface. For comparison, we calculate the free 2H-TPP molecule as having a dihedral angle of 62.7 degrees, in agreement with refs. [20, 21, 22, 10]. This distortion is also visible in the STM images in figure 1b. As a result of the rotation of the phenyl arms, the pyrrole rings containing the N-H motifs distort downwards and those containing the lone nitrogen atoms distort upwards. The composition of the molecular orbitals of the metal-adsorbate system was decomposed into contributions from occupied and unoccupied orbitals of the finite copper cluster and of the 2H-TPP. The resulting interaction diagram revealed a state at about 1.1 eV below the Fermi level, which contains character from the 2H-TPP HOMO (57%) and various copper slab MOs (43%) (see supplement), indicating a strong overlap and interaction (hybridization) between these two states. A similar calculation for Ag(111)-TPP showed a much weaker interaction between silver slab MOs (10%) and the 2H-TPP HOMO (90%). The distance between the silver slab and the 2H-TPP is large, so the resulting overlap between these two orbitals is small, and the bond is weak. Our findings are in agreement with the increase in nobleness of a metal descending down Group 11 from Cu to Ag to Au [2]. The interaction of the 2H-TPP HOMO with the Cu states results in a deep-lying filled bonding state, with almost equal contribution from the Cu and the molecule and a concomitant transfer of charge to the copper surface. The build of charge surrounding the copper-molecule interface (the red isovalue in figure 4c), along with the charge on the molecules themselves, both which are much larger for the 2H-TPP – Cu system than for 2H-TPP - Ag, prevents the adsorbate molecules from interacting with one another due to electrostatic repulsion, thereby impeding self-assembly on the Cu(111) surface.

The charge density difference plots in figure 4 reflect strong variations in the charge density at the interface upon adsorption: there is charge depletion directly under the center of the molecules and an increase in the charge density along the edges of the molecule and under the phenyl ligands. The underlying mechanism here is Pauli repulsion, which follows from the quantum mechanical requirement that overlapping electronic states must be orthogonal to each other. This drives up the energy and as a result pushes charge away at the surface of the Cu in an area directly under the center of the molecule. This effect has been described as the 'pillow effect' [23, 3, 24, 25]. The redistribution and exchange of charge also changes drastically the surface dipole of the Cu and is at the origin of the observed spatial variation of the work function. By comparison, the negligible charge transfer from Ag to the adsorbate is in line with a weaker binding energy, a longer metal-adsorbate distance and a negligible pillow effect occurring on the metal's surface.

Besides the electrostatic repulsion between the molecules there are also attractive interactions, mainly van-der-Waals and dispersive interactions. Additional bonding contributions come from CH-π and π–π interactions between the phenyl ligands. For a freestanding 2H-TPP dimer, the total binding energy was estimated to be 0.3 eV. The net effect is thus dependent on the competition between Coulomb repulsion and the mainly van der Waals attraction. The net force is attractive for 2H-TPP on Ag(111) and Au(111) [18, 10, 26] and repulsive on Cu(111), owing to the discussed differences in charge transferred and Coulomb repulsion. A similar dominance of the electrostatic repulsion has been reported earlier for other organic-metallic interface systems [27, 28, 10, 29, 30]. In addition here, the distorted phenyl arms of the molecules impede the formation of π–π bonds, thereby further decreasing the propensity of binding between two 2H-TPP molecules.

The observed differences in the interactions of 2H-TPP on Ag and Cu surfaces were exploited to actually control the inter-molecular forces, between the repulsive and attractive limits by engineering the metal-organic interface. The trick is to deposit the molecules on the Cu(111), which was pre-covered by an Ag buffer layer of variable thickness. The STM images of 2H-TPP adsorbed on 1 to 3 monolayers of Ag on Cu(111) are shown in figure 5. Clearly, the molecules remain, more or less, statistically distributed on 1 ML Ag/Cu, while islands of extended networks, identical in architecture to that found on Ag(111) in figure 1a, are observed for the 2H-TPP adsorbed on 3 ML Ag/Cu. At the intermediate Ag buffer layer thickness of 2 ML, clusters of 2H-TPP adsorbed molecules are commonly observed but with noticeable degree of disorder within such clusters. On Ag layers on Cu, the 2H-TPP molecules appear in 2 distinctively different symmetries: the symmetry labeled (i) which is usually observed on Cu(111), and the symmetry labeled (ii) which is typical for TPP on Ag(111). With increasing Ag layer thickness, the occurrence of 2H-TPP molecules in configurations of type (i) decreases while at the same time the occurrence of the 2H-TPP adsorbed molecules in the arrangement of type (ii) increases. It appears as if clusters of molecules, ordered or disordered, are mostly formed by 2H-TPP adsorbed molecules of type (ii).

Tunneling spectroscopy was again employed to elucidate the local electronic structure of the interface. While the spectrum of electronic states taken on top of the molecules does not show significant differences for all samples in figure 5, the Shockley surface state of the substrate on the other hand shifts upward in energy with increasing thickness of the Ag buffer layer, from -400 meV for clean Cu(111) to -50 meV for Ag(111) (figure 5d). The energy of this Shockley state is thus a precise indicator for the Ag layer thickness. Key is that the energy of the Shockley state can be adjusted by the Ag buffer layer thickness between the two extremes of pure Ag(111) and Cu(111) surfaces, and that has profound consequences for the molecular self-assembly as just demonstrated.

It is has been already established that the electronic level alignment at the metal-organic interface and frontier orbital symmetry determines the hybridization of levels and the amount of charge transferred across the interfaces [1, 31]. Based on the results shown here, we find it reasonable to assume that the Shockley state plays a crucial role for the interaction strength. Depending on the exact energetic position of this state, more or less overlap with the corresponding molecule levels is possible, thereby facilitating (Cu) or impeding (Ag) charge transfer across the interface. By controlling the exact energetic position of the surface state by the choice of thickness of Ag buffer layers on Cu(111), the degree of electronic level hybridization can thus be finely tuned, to adjust the amount of charge transferred and the strength of the Coulomb repulsion. Since the van der Waals interaction remains unaffected by this, the net effect can thus be chosen to be repulsive or attractive. The ability to control inter-molecular forces for a particular type of molecule between both extremes in this manner opens new possibilities to steer molecular self-assembly, especially if patterned buffer layers are used. It is thus an important new milestone in establishing rational design principles for organics in contact with surfaces. Specifically, we demonstrated the potential of using substrates to build organic structures and frameworks of potentially greater complexity than currently possible, exhibiting pre-defined and desired functionality.

**Methods**

The experiments were carried out using a Omicron low temperature scanning tunneling microscope in a ultrahigh vacuum system with a base pressure of 8 x $10^{-11}$ mbar. Single crystalline substrates have been cleaned in UHV by $Ar^+$ ion sputtering and annealing. TPP molecules have been deposited by thermal evaporation from a home-build Knudsen cell, with the substrate held at room temperature. The combined photoemission and inverse photoemission spectroscopy measurements have been performed in a second UHV system [32] but using the same substrates and Knudsen cells.

The DFT-D calculations were carried out using the ADF software package [33, 34]. The revPBE gradient density functional [35] was employed, and Grimme's latest dispersion corrected functional [36] was used to account for the dispersion forces. Tests were performed to determine the effect of the basis set on the binding energy of benzene to an Ag(111) slab. For the results given in the main text, the basis functions on all of the atoms consisted of a valence triple-ζ Slater-type basis set with polarization functions (TZP) from the ADF basis-set library. The core shells up to 1s, 1s, 3p and 4p of carbon, nitrogen, copper and silver, respectively, were kept frozen. In situations where SCF convergence issues arose, the steepest decent method was employed. A Mulliken charge analysis was used to determine the magnitude of the charge transferred between the adsorbate and the metal surface. More computational details and complementary results are provided in the supplementary material.

**Author contributions**

All authors contributed extensively to the work.

**Acknowledgements**

This work was supported by the National Science Foundation, in parts by the NSF grants DMR-0747704 (CAREER), DMR-0213808 (MRSEC) and CHE-0909580. Support from the Center of Computational Research at SUNY Buffalo is acknowledged.

**Additional Information**

The authors declare no competing financial interests.

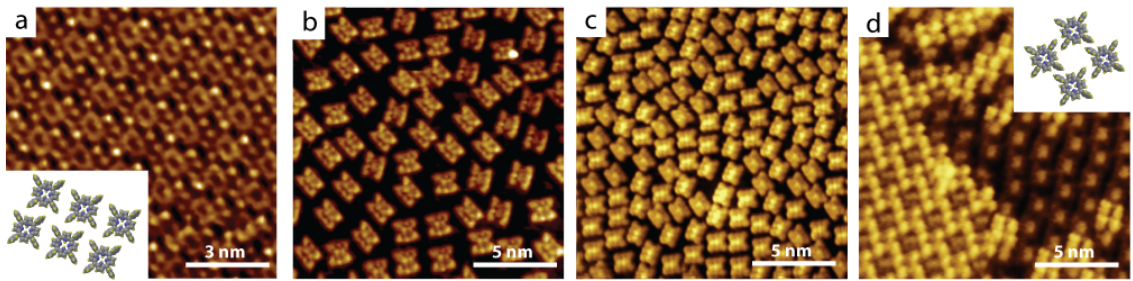

**Figure 1:** STM images of 2H-TPP on Ag(111) (a) and Cu(111) (b-d). The molecule coverage is 0.6 ML (b), 0.7 ML (c) and 1.2 ML (d). $I_t$ = 0.4 nA, $U_b$ = +0.8 V.

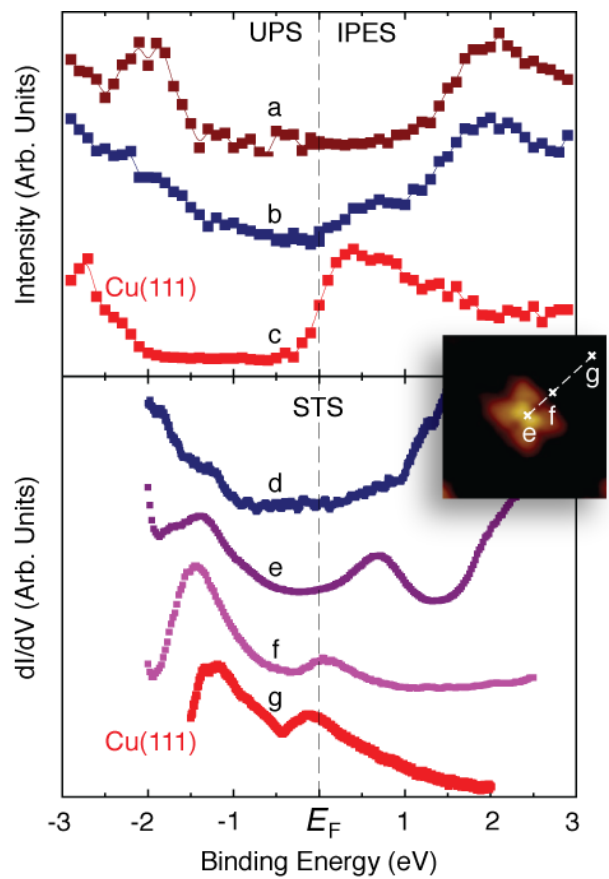

**Figure 2:** *Upper panel:* Photoemission (UPS) and inverse photoemission (IPES) spectra of 2H-TPP on Cu(111). (a) 3 ML 2H-TPP; (b) 1 ML 2H-TPP; (c) bare Cu(111). *Lower panel:* STS point spectra taken on or near lone TPP molecules. (d) on top of a molecule in the second layer; (e) on top of a molecule in the first layer; (f) on Cu, at a distance of 1 Angstrom from the molecule edge; (g) on Cu, several Angstroms away from a molecule. Inset: STM image showing the positions where spectra (e-g) were taken. Binding energies are denoted as $E-E_F$ making occupied state energies negative and unoccupied states positive.

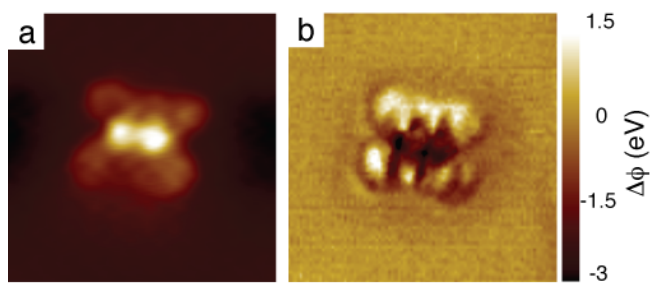

**Figure 3:** (a) STM image of a lone 2H-TPP molecule. $I_T$ = 0.4 nA, $U_b$ = +0.8 V; image size 4 nm x 4 nm. (b) Work function map of the same molecule, showing lowered work function at the center of the molecule and increased work function at the boundary of the molecule, relative to the substrate.

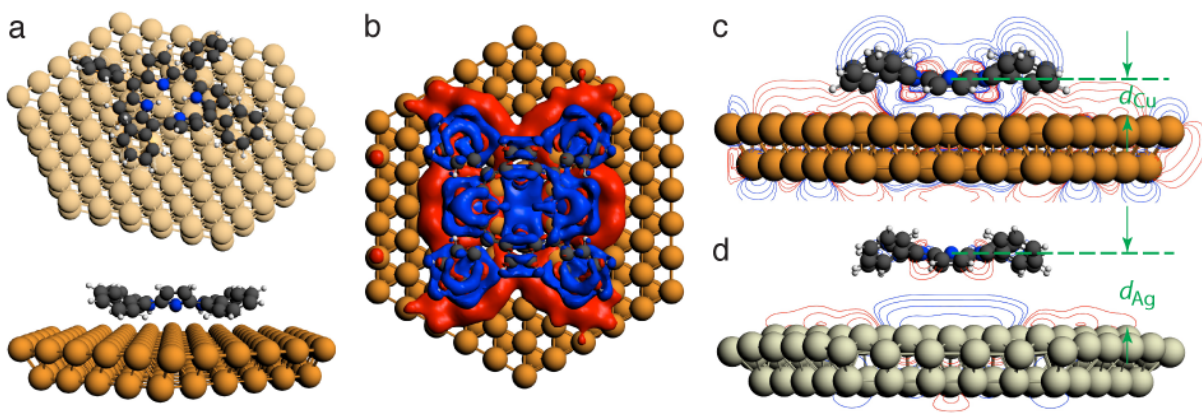

Figure 4: (a) Top and side view of the optimized geometry of 2H-TPP on top of a Cu(111) slab. (b, d) The calculated differences between the charge density of the metal-organic systems and that of the isolated, distorted fragments illustrates how the charge density changes upon adsorption of the molecule to the metal surface, with blue being a decrease and red an increase. (b) 2H-TPP on Cu(111) with an isovalue of +/- 0.0003 au. (c, d) Contour diagrams for 2H-TPP on top of Cu(111) (c) and Ag(111) (d). The same settings were employed to obtain the contours. The values $d_{Cu}$, $d_{Ag}$, are listed in Table 1.

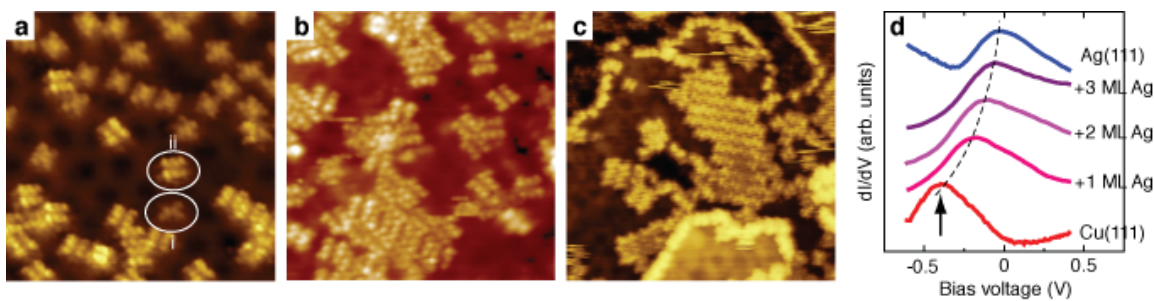

**Figure 5:** STM images of 2H-TPP on Ag buffer layers of thickness tAg on Cu(111) where tAg = 1 ML (a), tAg = 2 ML (b), tAg = 3 ML (c). Two different shapes of the molecules are observed, labeled (i) and (ii), see text for explanation. Image size 15 nm x 15 nm. (d) STS point spectra on Cu, Ag films on Cu, and Ag substrates showing the Shockley surface state.

| System | $E_b$ (eV) | Surface charge | Adsorbate charge | Dihedral angle (deg) | Surface distance $d$ (Å) |
|---|---|---|---|---|---|
| Cu(111) | 3.96 | -1.69 | 1.69 | 40.4 | 3.04 |
| Ag(111) | 0.42 | 0.02 | -0.02 | 49.2 | 7.02 |

**Table 1:** The binding energy ($E_b$), charge and structure of 2H-TPP on top of Cu(111) and Ag(111). The surface distances, $d_{Cu}$ and $d_{Ag}$, are the average calculated distances of the central nitrogen atoms of the 2H-TPP to the metal. The dihedral angle is defined in the supplemental material.